\documentclass{article}

\usepackage{microtype}
\usepackage{graphicx}
\usepackage{subfigure}
\usepackage{booktabs} % for professional tables
\usepackage{csquotes}
\usepackage{todonotes}

\usepackage{hyperref}

\usepackage[accepted]{icml2021}

\icmltitlerunning{Decision Tree Learning in K-12 Education}

\begin{document}

\twocolumn[
\icmltitle{Data, Trees, and Forests -- Decision Tree Learning in K-12 Education}



\icmlsetsymbol{equal}{*}

\begin{icmlauthorlist}
\icmlauthor{Tilman Michaeli}{tum}
\icmlauthor{Stefan Seegerer}{fu}
\icmlauthor{Lennard Kerber}{on}
\icmlauthor{Ralf Romeike}{fu}

\end{icmlauthorlist}

\icmlaffiliation{tum}{Computing Education Research Group, Technical University of Munich, TUM School of Social Sciences and Technology}
\icmlaffiliation{fu}{Computing Education Research Group, Freie Universität Berlin}
\icmlaffiliation{on}{Otto-Nagel-Gymnasium, Schulstr. 11, 12683 Berlin}

\icmlcorrespondingauthor{}{tilman.michaeli@tum.de}
%\icmlcorrespondingauthor{Eee Pppp}{ep@eden.co.uk}

\icmlkeywords{Machine Learning, Computing Education, K-12, Supervised Learning, Data Literacy, Decision Trees, CS Unplugged}

\vskip 0.3in
]



\printAffiliationsAndNotice{}  % leave blank if no need to mention equal contribution

\begin{abstract}
As a consequence of the increasing influence of machine learning on our lives, everyone needs competencies to understand corresponding phenomena, but also to get involved in shaping our world and making informed decisions regarding the influences on our society. Therefore, in K-12 education, students need to learn about core ideas and principles of machine learning. However, for this target group, achieving all of the aforementioned goals presents an enormous challenge. To this end, we present a teaching concept that combines a playful and accessible unplugged approach focusing on conceptual understanding with empowering students to actively apply machine learning methods and reflect their influence on society, building upon decision tree learning.
\end{abstract}

\section{Introduction}

Due to rapid technological progress, more and more areas of our lives are influenced and shaped by machine learning (ML). In consequence, everyone needs the competencies to be able to adequately analyze, discuss and help shape the impact, opportunities, and limits of artificial intelligence on their personal lives and our society. To this end, beginning as early as in K-12 education, students need to learn about core ideas and principles of AI and machine learning in particular \cite{wcce-ai}. This is reflected in an increasing number of K-12 computer science curricula being extended to include the topic of AI \cite{aaai, yu2018design}. From the students' point of view, exciting questions arise here, such as whether machines can \enquote{think}, how they \enquote{learn}, or what \enquote{intelligence} lies beyond everyday phenomena such as parking aids, image recognition, or purchase recommendations in online stores.

Closely linked to the topic of ML is data science, which is of particular interest for approaches in machine learning and thus also reflected in multiple AI curricula. 
Corresponding methods are also used to gain knowledge in a wide variety of scientific disciplines. Data analysis and artificial intelligence are often referred to as the fourth pillar of science \cite{riedel2008classification, tolle2011fourth}. This is becoming increasingly relevant for K-12 education as well, as this shift in the scientific disciplines is also reflected in corresponding subjects. 
This results in challenges for education: it is important to not only make it possible for students to explain AI-related phenomena in their daily lives, but also to enable them to actively and creatively shape the so-called \enquote{digital} world. 
Therefore, in this paper we report on a teaching concept for the K-12 classroom, focusing on supervised learning with decisions trees, empowering students in a playful manner to understand and actively use ML methods even with no prior knowledge.

\section{Related Work}
Decision Tree Learning (DTL) is considered a rather simple and graspable (yet relevant in real applications) ML method \cite{lin2013data} that exemplifies the idea of supervised learning. Therefore, it is often used in curricula \cite{touretzky2021artificial, heinemann2018drafting} or teaching materials. With regards to K-12 education, we see numerous unplugged approaches for introducing DTL, such as \emph{AI Unplugged} \cite{lindner2019unplugged} or using \emph{data cards} \cite{10.1145/3488042.3489966}. Approaches that work without a computer are particularly suitable for demonstrating CS concepts in a fun and engaging way \cite{bell2009computer}. However, they do not allow for actually using those concepts to creatively design computational artifacts. The same goes for other web applications that aim for demonstrating and experimenting with DTL, such as \cite{elia2021interactive} or \cite{mariescu2019machine}. 
In contrast, approaches that build upon programming such as \cite{biehler2021introducing} enable using DTL creatively and in real-world scenarios. However, the necessary programming skills provide a significant barrier in secondary education. Given the manifold problems and complexity of learning programming \cite{robins201912}, this can draw attention away from the actual learning objective of ML. To address this challenge, we present a teaching concept combining unplugged activities and Orange\footnote{https://orangedatamining.com}, a visual data analysis tool commonly used within computing education in the context of data mining \cite{grillenberger2019classes}. This approach allows for understanding the principles of DTL and supervised learning in a low-barrier, engaging way, as well as actively working with respective methods on realistic data sets.

\section{Teaching Concept Outline}
The teaching concept we describe in the following is aimed at students from age 14. Due to extensive support materials, no prior computer science or ML knowledge is required for teachers or students - although both would be beneficial, as it allows for deeper coverage of certain topics. The concept\footnote{All materials   are available under a creative commons license here: https://computingeducation.de/proj-it2school/} (see table \ref{concept}) is divided into four phases: (1) Firstly, using an unplugged activity, the students develop and test a decision tree for a given data set manually. (2) Building upon the previous step, they train a model using Orange for the same data set and compare the \enquote{manual} and \enquote{automated} procedures. (3) This is followed by applying the concepts learned to real data sets in a project phase. (4) Building on this application of AI methods, the social dimension is then considered, and pressing questions regarding transparency, fairness, and security of AI processes are discussed.

\begin{table}[h]
\caption{Outline of the teaching concept.}
\label{concept}
\vskip 0.15in
\begin{center}
\begin{small}
\begin{sc}
\begin{tabular}{p{.15\columnwidth}p{.65\columnwidth}}
\toprule
\#Lessons (45 min) & Phase  \\
\midrule
1 & Unplugged Decision Tree Learning \\
2 & Decision Tree Learning in Orange \\
2-3 & Project Phase \\
1 & AI and Society \\
\bottomrule
\end{tabular}
\end{sc}
\end{small}
\end{center}
\vskip -0.1in
\end{table}

Orange \cite{demvsar2013orange} is a data analysis tool developed at the University of Ljubljana primarily aiming at educational contexts and available for download for all common desktop operating systems, without requiring any license. In Orange, data analysis is described as a data flow diagram rather than text-based programming (see figure \ref{Orange}). Accordingly, no syntax errors occur and teaching can focus on the analysis and interpretation of the data. 

With this \enquote{data-oriented} approach, this teaching concept allows learners without programming experience to apply and explore complex AI methods and discover important, current issues of societal scope themselves, and discuss them based on realistic scenarios. By taking a data-oriented view, there are possible applications or connections for a variety of subjects, for example in the natural sciences or economics. There is also the opportunity to grasp the causes and opportunities of the profound changes resulting from digital transformation. In doing so, students not only apply AI methods but also learn about career prospects such as that of data scientists and discuss the societal impact that would result from the use of their models. Hence, the learning objectives address the range from understanding the functionality of AI systems (e.g. describing how the training examples provided in an initial data set can affect the results of an algorithm), to applying AI methods in a constructive manner (e.g. applying the steps of ML to solve a specific data-based problem), and to thinking about possible effects of using AI systems (e.g. ...explaining why biases in data affect the results of machine learning and discussing implications for the use of AI systems) \cite{wcce-ai}. In the following, we describe the four phases of the teaching concept in detail. 

% \todo[inline]{ggf. Lernziele}

\begin{figure}[h]
\vskip 0.2in
\begin{center}
\centerline{\includegraphics[width=.9\columnwidth]{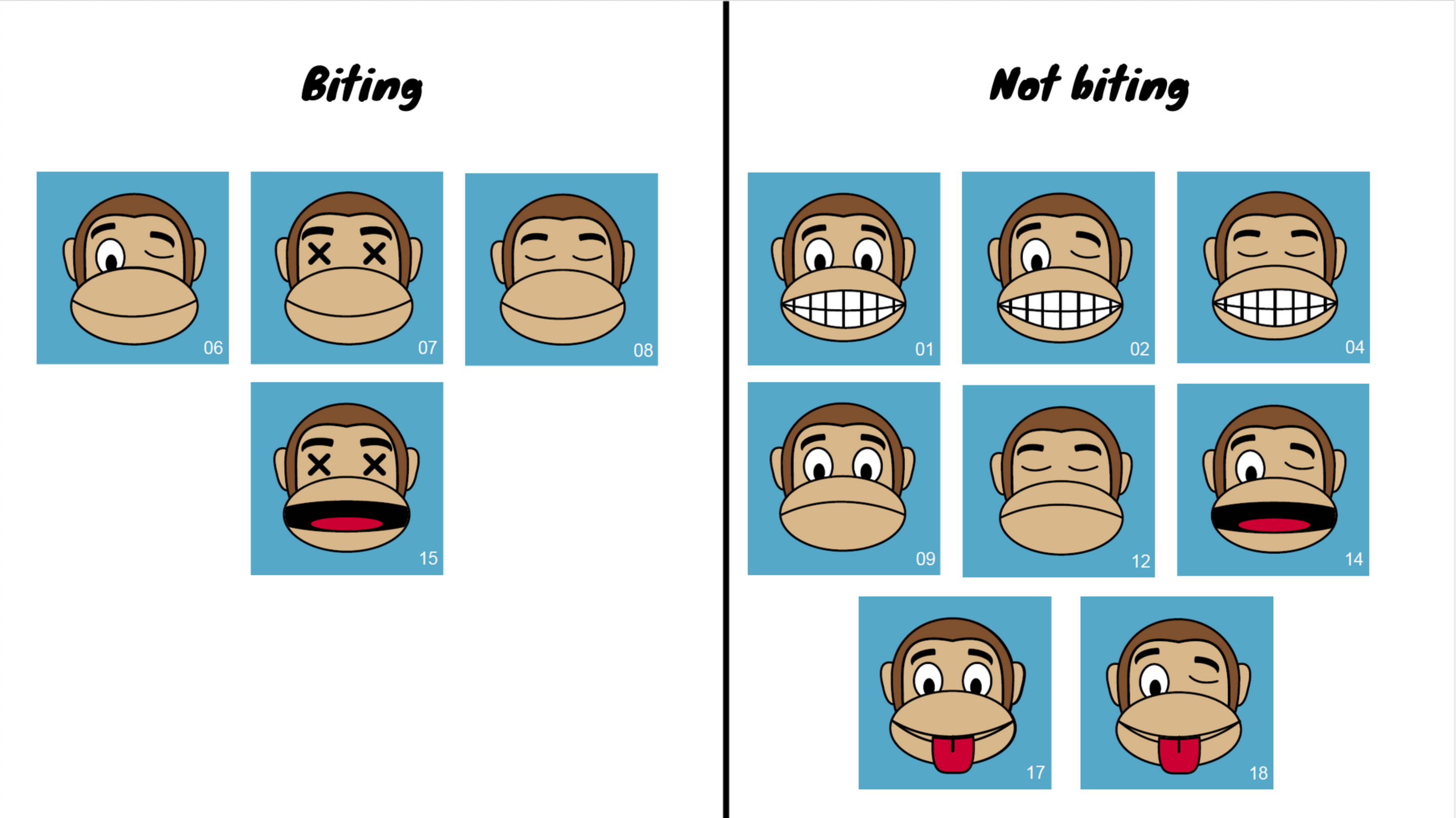}}
\caption{Data set \enquote{monkeys}}
\label{monkeys}
\end{center}
\vskip -0.2in
\end{figure}

\subsection{Unplugged Supervised Learning with Decision Trees}
\begin{figure}[h]
\vskip 0.2in
\begin{center}
\centerline{\includegraphics[width=\columnwidth]{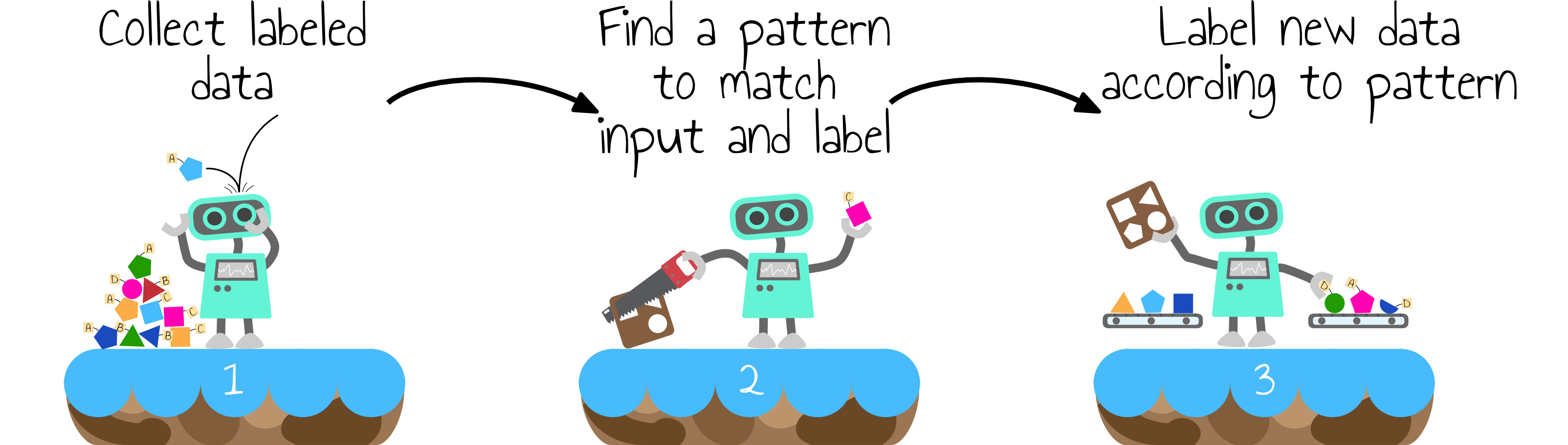}}
\caption{Supervised Learning}
\label{SL}
\end{center}
\vskip -0.2in
\end{figure}

In the first phase, students learn about the concept of DTL and confusion matrices, as well as metrics for assessing the quality of a model, based on Activity 1 of AI Unplugged \cite{lindner2019unplugged}. After reflecting on a newspaper story that demonstrates the power of data analysis, they are confronted with a set of biting and non-biting monkeys (see figure \ref{monkeys}). As zookeepers, they are tasked with deriving rules to distinguish between those two classes of monkeys based upon the monkeys' features, such as \enquote{IF open eyes THEN does not bite}. Afterward, the idea of representing those rules as a decision tree is introduced, and the students manually generate a decision tree for a more complex monkey data set. Based on test data, students investigate the quality of their respective models. To this end, they produce a confusion matrix and calculate metrics such as accuracy, precision, and recall. This is followed by highlighting the underlying concept of supervised learning (see figure \ref{SL}), as the students describe which steps were carried out one after the other.

\subsection{Decision Tree Learning in Orange}
\begin{figure}[h]
\vskip 0.2in
\begin{center}
\centerline{\includegraphics[width=.75\columnwidth]{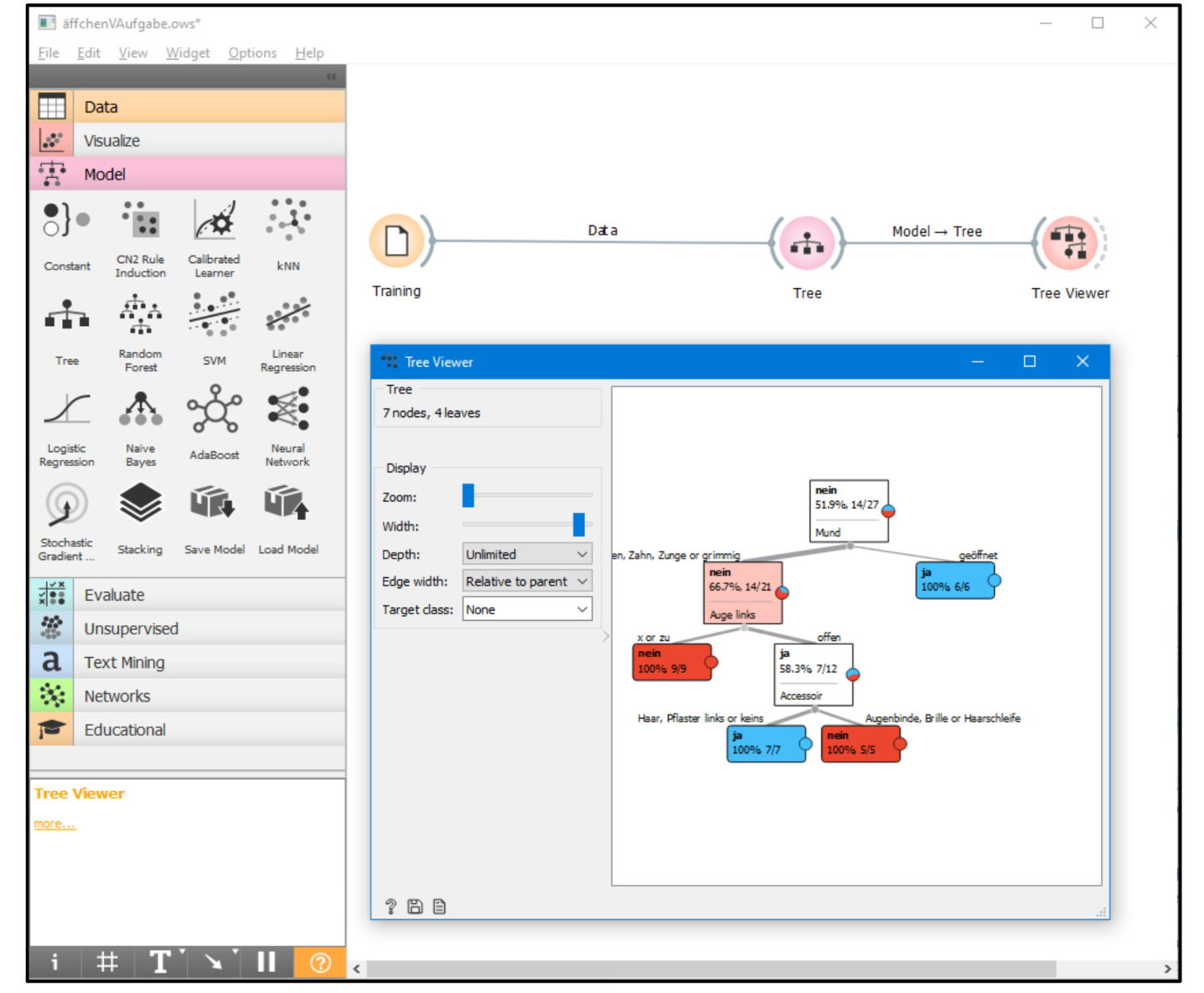}}
\caption{Decision Tree Learning in Orange}
\label{Orange}
\end{center}
\vskip -0.2in
\end{figure}

In the second part, students learn to use Orange as a tool to analyze given data to be able to carry out the project. To this end, the students firstly model the monkey data set as a table, so that a computer can process them. There are different possibilities for modeling the data set, e.g. binary (feature present/not present), or via the different feature characteristics (e.g. mouth: open/closed/tongue out/teeth bared). Afterward, the students use Orange to train a model for distinguishing biting and non-biting monkeys via DTL (see figure \ref{Orange}). 

Furthermore, the students use Orange to compute a confusion matrix and respective metrics (accuracy, precision, recall) regarding the quality of their model. They compare their \enquote{manual} process and results within the unplugged activity with this \enquote{automated} approach. Additionally, they experiment with different hyper parameters to explore over- and underfitting, in particular.

\begin{figure}[h]
\vskip 0.2in
\begin{center}
\centerline{\includegraphics[width=.95\columnwidth]{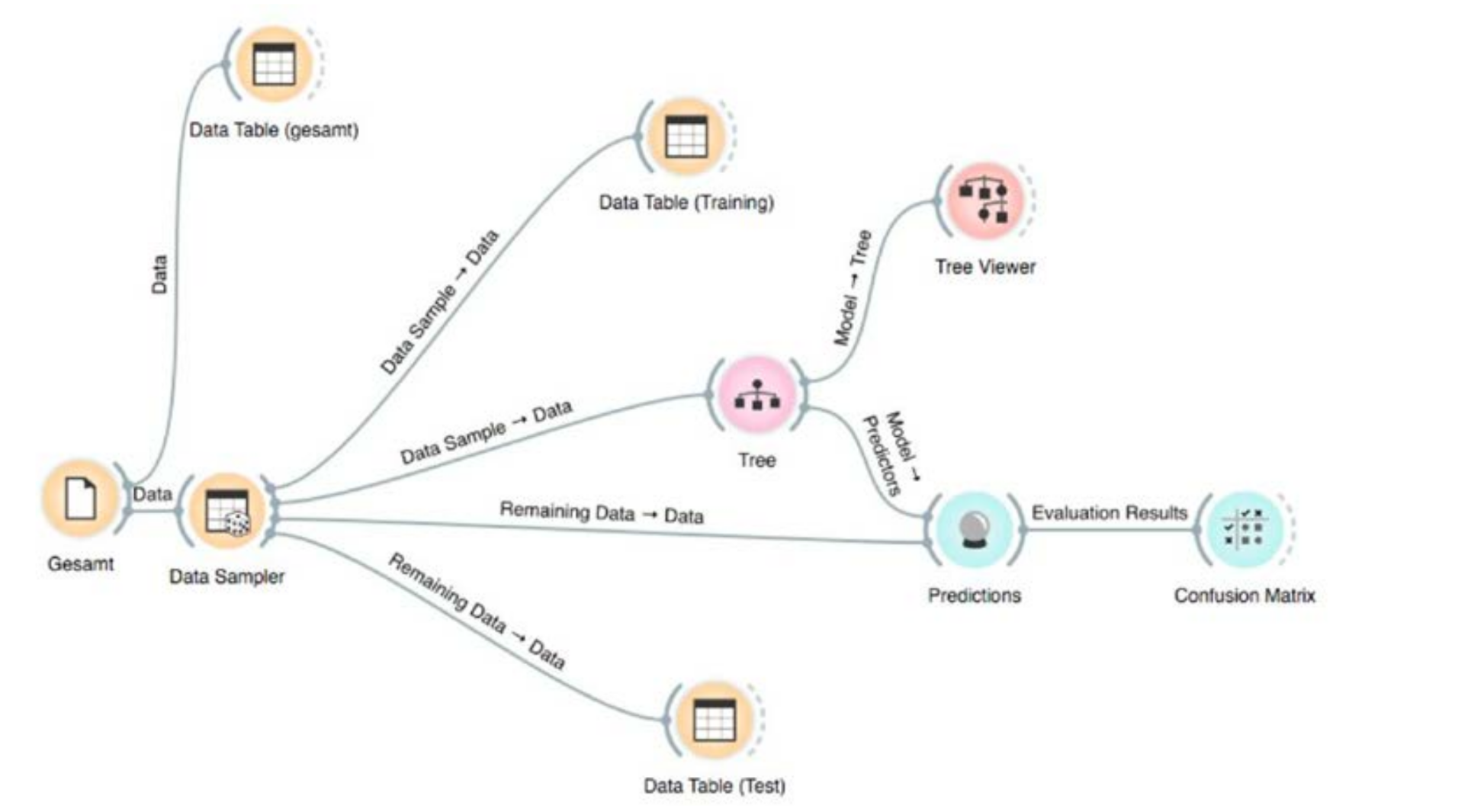}}
\caption{Exemplary Project}
\label{Orange2}
\end{center}
\vskip -0.2in
\end{figure}

\subsection{Project Phase}

During the two- to three-hour project work phase, students work largely independently. Firstly, different topics and related questions for the project work are presented to the class, groups are formed and choose a topic each. Exemplary data sets provided are predicting employee turnover based on employee data, predicting potential customers for fixed deposits based on customer profiles, predicting whether a class will be passed based on student data, predicting breast cancer based on measured cell nucleus features, and so on. Similar data sets have been chosen in a curriculum for data analysis by \cite{grillenberger2019classes}.
    
The students explore the respective data set, available features, and classes. Afterward, they split their data set into training and test data, train multiple models using different hyper parameters and features, and test and evaluate their models using suitable metrics for the respective use case. Furthermore, students are tasked with characterizing the societal implications of their model, such as defining persons that would profit or could be disadvantaged by using the model in real life. At the end of the project phase, the students present their results by creating posters for a gallery walk. For more advanced students, further supervised learning methods for classification such as random forests or k-nearest-neighbor can be introduced and used in the projects.

\subsection{AI and Society}
In the last part, the societal and ethical perspective of using AI models is discussed further, based on the knowledge about the chances and limits of ML methods that students experienced first-hand in their project work. For this purpose, students are presented with three scenarios (applicant selection, facial recognition, automated grading in the UK) of unreliable, discriminatory, or faulty AI systems. Students answer questions such as \enquote{What could be the cause(s) of the problematic decisions made by the AI systems in the scenarios described?}, \enquote{Based on your experience with training models: How could these problems be solved} or \enquote{What rules should apply to fair use of AI systems?}. Using a think-pair-share approach, the questions are first discussed in partner work and then in plenary.

\section{Experiences from the Classroom}
We conducted the teaching concept with year 10 students with limited prior formal CS experience. In the classroom, we noted that many students were not aware of the difference between correlation and causation, which is crucial for assessing the capabilities and limits of ML systems -- potentially as they are used to think in cause-effect-relations. 
The unplugged activity served as a low-barrier entry to the topic and helped the students to convey and transfer the idea towards analyzing the data using Orange. 
During the project phase, the students succeeded in creating various models by applying DTL on the data sets. This allowed them to experience the basic procedure of training a model and configuring hyper parameters with subsequent testing for themselves. The simple usage of Orange was very helpful here. Furthermore, students became aware that supervised learning is typically divided into a training and testing phase, and repeatedly mentioned that a model does not produce completely correct results in every case. Therefore, they developed a basic understanding of the opportunities and the limits of ML. Furthermore, with this approach, we were able to successfully convey not only DTL as a particular ML method, but also the underlying concept and principles of supervised learning. 

With regards to the data sets they worked with, students were surprised by some of the features provided, such as the attribute race in an American census data set. Another group working on the topic of passing a class was irritated by the possibility of integrating supposedly extraneous features for the prediction, such as the educational level of the parents. We conclude that this way students became aware of ethical and societal concerns in ML. Overall, for the students, it was eye-opening to experience how easy it is to analyze such large amounts of data.

\section{Discussion and Conclusion}
With the teaching concept outlined, we combine the potential of unplugged and design-oriented approaches to address the challenge of teaching ML in secondary education: The unplugged monkey game allows for a low barrier, hands-on and playful look into the black box that makes the concept of DTL graspable (independently of the specific implementation). Furthermore, students learn about underlying concepts of supervised learning, such as training and testing, evaluating the quality of a model, the difference between correlation and causation, and corresponding consequences on the quality of predictions. This focus on underlying ideas and principles ensures the long-term relevance of the content for the students' future lives, in contrast to short-lived technological details and developments \cite{denning2004great} and illustrates how DTL is suitable for exemplifying supervised learning.  

Building upon this, students apply these concepts to realistic data sets, as Orange enables them to avoid the otherwise necessary and (for secondary education) complex programming skills. This way, they can compare their \enquote{manual} and the \enquote{automated} generation of decision trees, can experiment and explore with the data set and hyper parameters, and discover that even professionals need some experience in choosing them adequately. While working in Orange, the unplugged activity helps students to reflect on underlying concepts. In the project phase, students work actively and creatively with realistic data sets in a constructionist way, designing their own computational artifacts. This experience enables them to discuss the potentials, limits, and important ethical considerations of ML in our society on a sound foundation. This is also reflected in our experience from the classroom, where students were stunned by how easily large amounts of data can be analyzed and which seemingly unrelated or even problematic features might be used for training models. 

Furthermore, the skills conveyed in this teaching concept are becoming increasingly relevant beyond computing education. In the context of digital transformation and its impact on all school subjects and their related scientific disciplines, everyone needs competencies in handling and analyzing data. Therefore, the concept offers an enormous potential for action-oriented and interdisciplinary teaching in the K-12 classroom. Whether conducting experiments in physics classes, answering questions in chemistry or biology classes, evaluating geodata in geography classes, analyzing text documents in humanities, or discussing ethical, social, and moral consequences in ethics or religion classes: the methods introduced in this teaching concept can be used across disciplines to answer subject-specific questions. To this end, we are gathering more data sets that particularly enable this interdisciplinary perspective. Furthermore, we are currently exploring this teaching concept in tertiary education to provide non-CS students with the necessary skills in the context of ML and data science to answer subject-specific scientific questions -- especially as no- and low-code approaches are gaining popularity in the professional world \cite{sanchis2019low}.

In summary, this teaching concept allows for an understanding of how supervised learning and, in particular, DTL works, how it can be applied efficiently to solve problems, and how it might interact with individuals and society \cite{dagstuhl}. In consequence, it empowers students even with no or little CS knowledge to understand AI phenomena surrounding them and actively and creatively shape the \enquote{digital world} and hence may serve as good practice example in AI education.

\bibliography{main.bib}
\bibliographystyle{icml2021}

\end{document}